\documentclass[aip,amsmath,amssymb, reprint,cha]{revtex4-1}
\usepackage[utf8]{inputenc}
\usepackage[T1]{fontenc}
\usepackage{graphicx}
\usepackage{bm}
\usepackage{amssymb}
\usepackage{amsmath}
\usepackage{times}
\usepackage{color}
 \usepackage[hidelinks,unicode=true]{hyperref}
\usepackage[]{hyperref}
\hypersetup{colorlinks=true,linkcolor=blue,urlcolor=blue,citecolor=blue,pdfhighlight=/N}
\usepackage{lipsum}
\usepackage{autonum} 
\usepackage{enumitem}

\def \dd {{\rm d}}

\DeclareMathOperator{\Real}{Re}

\graphicspath{{figures/}}

\makeatletter
\def\@email#1#2{%
 \endgroup
 \patchcmd{\titleblock@produce}
  {\frontmatter@RRAPformat}
  {\frontmatter@RRAPformat{\produce@RRAP{*#1\href{mailto:#2}{#2}}}\frontmatter@RRAPformat}
  {}{}
}%
\makeatother
\begin{document}

\preprint{AIP/123-QED}

\title[]{From chimeras to extensive chaos in networks of heterogeneous Kuramoto oscillator populations}

\author{Pol Floriach}
\affiliation{ 
Department of Mathematical Sciences, University of Copenhagen, Copenhagen (Denmark)
}

\author{Jordi Garcia-Ojalvo}
\affiliation{ 
Department of Medicine and Life Sciences, Universitat Pompeu Fabra, Barcelona (Spain)
}

\author{Pau Clusella}
 \email{pau.clusella (AT) upc.edu}
\affiliation{ 
EPSEM, Departament de Matemàtiques, Universitat Politècnica de Catalunya, Manresa (Spain)
}%

\date{\today}
\begin{abstract}

Populations of coupled oscillators can exhibit a wide range of complex dynamical behavior, from complete synchronization to chimera and chaotic states.
We can thus expect complex dynamics to arise in \textit{networks} of such populations. Here we analyze the dynamics of networks of populations of heterogeneous mean-field coupled Kuramoto-Sakaguchi oscillators, 
and show that the instability that leads to chimera states in a simple two-population model also leads to extensive chaos in large networks of coupled populations. 
Formally, the system consists of a complex network of oscillator populations whose mesoscopic behavior evolves according to the Ott-Antonsen equations.
By considering identical parameters across populations, the system contains a manifold of homogeneous solutions where all populations behave identically.
Stability analysis of these homogeneous states provided by the master stability function formalism shows that non-trivial dynamics might emerge on a wide region of the parameter space for arbitrary network topologies.
As examples, we first revisit the two-population case, and provide a complete bifurcation diagram. Then, we investigate the emergent dynamics in large ring and Erdös-Rényi networks. In both cases, transverse instabilities lead to extensive space-time chaos, i.e., irregular regimes whose complexity scales linearly with the system size. 
Our work provides a unified analytical framework to understand the emergent dynamics of networks of oscillator populations,
from chimera states to robust high-dimensional chaos.

\end{abstract}
\maketitle

\begin{quotation}
Many physical systems are composed of large numbers of interacting oscillatory units.
Such complex systems typically display synchronization patterns that vary in complexity, depending on the oscillators' intrinsic dynamics and the nature of their interactions.
For instance, the Kuramoto-Sakaguchi model provides a simple framework in which a population
of globally coupled phase oscillators transitions from incoherence to synchronization as the strength
of the coupling is varied.
Interestingly, when coupling two populations of Kuramoto-Sakaguchi oscillators 
other collective dynamics may emerge, in which one population displays a higher degree of synchronization
than the other.
These regimes are an example of \emph{chimera states}, a phenomena that has been studied in 
a number of theoretical setups and experimental works. 
Here, we extend the analysis of these symmetry-breaking transitions to the case of $N$ 
oscillator populations interacting through arbitrary complex networks. 
In particular, for ring and random topologies, we reveal the existence of highly chaotic states 
whose complexity increases linearly with the system size, a situation known as \emph{extensive chaos}.
\end{quotation}

\section{Introduction}

Chimera states are an intriguing example of non-trivial collective activity emerging in networks of coupled oscillators.
In their simple original form, chimera states consist of a population of identical oscillatory units spontaneously
splitting into two subgroups: a completely synchronized cluster and a partially synchronized one~\cite{kuramoto2002,abrams2004}. 
This form of spontaneous symmetry breaking
was first uncovered in non-locally coupled rings of phase oscillators, and has since been studied in a wide variety of theoretical and real systems~\cite{parastesh2021,Haugland_2021}.

The Kuramoto-Sakaguchi model\cite{sakaguchi1986} provides a simple yet powerful
framework to understand the collective behavior of complex oscillatory systems, including chimera states.
In this model, each oscillator
corresponds to a single phase variable that interacts with the other units through an all-to-all sinusoidal coupling.
Remarkably, this mean-field setup
allows an exact low-dimensional reduction through the Ott-Antonsen ansatz\cite{ott2008,bick2020}.
This method provides
the collective dynamics of the system as a single differential equation for the complex Kuramoto order parameter
\footnote{See also
the Watanabe-Strogatz theory\cite{wang1996}, which provides a complete description for the homogeneous case.
For a comprehensive general theory that includes and relates Watanabe-Strogatz and Ott-Antonsen frameworks see Ref.~[\onlinecite{cestnik2022}].
}.

Researchers have employed the Ott-Antonsen equations to unveil the nature of chimera states
in systems composed of two identical populations of Kuramoto-Sakaguchi oscillators\cite{abrams2008,laing2009,laing2012,kotwal2017,guo2021}.
For oscillators with equal natural frequencies, the chimera state always coexists with the homogeneous solution (i.e., full synchronization of the two populations)\cite{abrams2008}.
However, for oscillators with distributed frequencies, the chimera state might emerge through a pitchfork bifurcation
of the homogeneous solution, thus attracting solutions are always inhomogeneous in wide regions of the parameter space\cite{laing2009,kotwal2017}.
Interestingly, chimera states are not always stable in these regions, leading to the emergence of other regimes
including symmetric oscillations, antiphase states, or even low-dimensional chaos\cite{laing2009,laing2012,kotwal2017,guo2021}
\footnote{For studies of non-identical populations (i.e., ensembles governed by different parameter values) see Refs.[\onlinecite{montbrio2004,martens2016,choe2016}].}.

A natural question to ask is how do these results transduce to large networks of more than two populations.
A number of recent studies have addressed this topic\cite{martens2010,martens2010a,lee2023a,lee2023,laing2023} (see also [\onlinecite{bick2020}] for a thorough review).
In particular, Martens\cite{martens2010,martens2010a} studied networks of three populations
without disorder,
and unveiled the emergence of different types of chimera states.
Lee and Krischer~\cite{lee2023a} extended these studies and showed the emergence of low-dimensional chaos in the three-population model.
The same authors also investigated six populations on a ring topology, and showed that chimeras are generally unstable,
but connected through attracting heteroclinic cycles\cite{lee2023}.
On the other hand, Laing~\cite{laing2023} investigated the emergence of chimera states in
rings composed of different number of populations with non-local coupling.
Interestingly, this work unveils chaotic dynamics appearing through the instability of chimera states, 
and shows that this irregular regime vanishes as the number of populations in the network increases.

In spite of all this progress, the behavior of identical populations of Kuramoto-Sakaguchi populations
interacting through complex networks
remains an open problem, due to the variety of possible topological configurations. 
In this paper we take on this challenge, and provide a general unifying framework to study the emergence of spatiotemporal
dynamics in these models.
We consider networks where the overall external input to each population is normalized across nodes,
thus ensuring the existence of homogeneous states.
The stability of these trivial regimes to small perturbations can be studied by means of
the master stability function\cite{pecora1998}, which provides the growth rate of arbitrary perturbations
as a function of the eigenmodes of the structural connectivity matrix.
This method allows to obtain an analytical expression for the dispersion relation of the homogeneous states.

We show that, for populations with intrinsic disorder, transverse instabilities
arise in a wide region of the parameter space for arbitrary network topologies.
In the case of two populations, such transverse instability corresponds to the pitchfork bifurcation known to lead to chimera states\cite{laing2009,kotwal2017}, whereas in large networks of ring and Erdős-Rényi topologies,
the dynamics becomes immediately chaotic with several positive Lyapunov exponents.
Numerical simulations reveal that this is a case of extensive chaos, a dynamical regime in which
 the dimension of the chaotic attractor increases linearly with the system size.
Extensive chaos, first conjectured by Ruelle\cite{ruelle1982},
has been widely studied in spatially extended systems (usually under the more general term of "space-time chaos")\cite{manneville1985liapounov,livi1986,keefe1989,ohern1996,egolf2000,xi2000,paul2007}, 
 networks of spiking neurons\cite{monteforte2010,palmigiano2022}, and mean-field models\cite{takeuchi2011}, among other systems.
Nonetheless, to the best of our knowledge, this phenomenon had never been reported in networks of Kuramoto oscillators before.

Overall, our study presents a general analytical approach to study the dynamics
of networks of identical Kuramoto-Sakaguchi populations,
and provides a direct link between the emergence of chimera states in two-population models and the appearance of extensive chaos in large networks of such populations.

\section{A network of Kuramoto-Sakaguchi oscillator populations}

Let us consider $N$ populations composed of $M$ Kuramoto-Sakaguchi oscillators each~\cite{sakaguchi1986,bick2020}.
Each oscillator in a population is described by their phase variable $\phi_{j,\sigma}$ where $j\in\{1,\dots,M\}$
and $\sigma \in\{ 1,\dots,N\}$.
The phase dynamics are given by
\begin{equation}\label{eq:general_kuramoto}
    \dot \phi_{j,\sigma} = \omega_{j,\sigma} + \frac{K}{M}\sum_{\kappa=1}^N c_{\sigma \kappa}\sum_{m=1}^M \sin\left(\phi_{m,\kappa}-\phi_{j,\sigma} - \alpha\right)
\end{equation}
where $\omega_{j,\sigma}$ are the natural frequencies of each oscillator,
$K$ is a global coupling strength parameter,
and $\alpha$ is the Sakaguchi phase-shift parameter.
In this equation, both the intra and inter-population couplings are set via a mean-field, i.e.,
if populations $\sigma$ and $\kappa$ are coupled,
then all oscillators in both populations contribute identically to the interaction.
Therefore the only source of irregularity in the interactions is set by the inter-population
connectivity, which is determined by a weighted complex network with connectivity matrix $C=(c_{\sigma \kappa})$.
We impose two constrains on this connectivity:
\begin{itemize}
    \item We assume that $C$ is row-normalized, i.e.,
$\sum_{\kappa=1}^N c_{\sigma \kappa}=1\;,$ thus, the total influence received by each oscillator is the same across populations. 
\item We assume that $C$ is diagonalizable. This condition is always fulfilled if the network is undirected.
\end{itemize}
Generally, we focus on cases where the internal coupling within each population
is larger than the coupling between other populations, i.e., $c_{\sigma \sigma} > c_{\sigma \kappa}$ for all $\sigma \neq \kappa$.
Nonetheless, this is not a strict requirement.

The natural frequencies of the oscillators $\omega_{j,\sigma}$ are distributed according to a Cauchy (Lorenz)
distribution centered at $\overline \omega$ and with half-width at half-maximum $\Delta$.
Due to the rotational symmetry of the model, we can set  $\overline \omega = \tan(\alpha)(K\cos(\alpha)-\Delta)$
without loss of generality, a choice that will be justified in the next section.

The complex Kuramoto order parameter of each population reads
\begin{equation}
Z_\sigma = R_\sigma e^{i\Phi_\sigma}=\frac{1}{M}\sum_{m=1}^M e^{i\phi_{m,\sigma}} \;.
\end{equation}
By considering now that $M\to\infty$,  we can apply the Ott-Antonsen ansatz~\cite{ott2008}, 
to obtain the evolution of $Z_\sigma$: 
\begin{equation}\label{eq:oaM}
    \dot Z_\sigma = (-\Delta + i\overline \omega)Z_\sigma + 
    \frac{K}{2}\sum_{\kappa=1}^N c_{\sigma \kappa} \left( Z_\kappa e^{-i\alpha}  - Z^2_\sigma Z^*_\kappa e^{i\alpha} \right).
\end{equation}
This equation provides the mean-field dynamics of each population in the network assuming the
phases follow a circular Cauchy distribution (Poisson kernel).
Moreover, it has been recently shown that the this ansatz is a global 
attractor of the Kuramoto-Sakaguchi model for $\Delta >0$\cite{cestnik2022}.
Therefore, Eq.~\eqref{eq:oaM} is going to be our main object of study for the rest of the article.

\section{Analysis of homogeneous states and their stability}\label{section:homogeneous}

In order to explore the dynamics of Eq.~\eqref{eq:oaM}, first we study its homogeneous manifold,
i.e., all the states in which the $N$ populations evolve with exactly the same dynamics.
By imposing $Z_\sigma=Z$ $\forall \sigma$ into Eq.~\eqref{eq:oaM}, and exploiting the fact that $C$ is row-normalized, we obtain
\begin{equation}\label{eq:hommanifold}
    \dot Z=  (-\Delta + i\overline \omega)Z +
    \frac{K}{2}\left( Z e^{-i\alpha}  - Z^2 Z^* e^{i\alpha} \right)\;,
\end{equation}
which corresponds to the mean-field equation for a single, isolated population of Kuramoto-Sakaguchi oscillators~\cite{ott2008,bick2020}.
The dynamics of this system are well studied: if $K<K_c := \frac{2\Delta}{\cos(\alpha)}$
the system converges to the incoherent state $Z:=0$. 
At $K=K_c$ the system undergoes a supercritical Hopf bifurcation (H$_0$ in the figures)
giving rise to a synchronized or coherent state, which takes the form $Z:= Re^{i\Omega t}$, with
\begin{equation}
\begin{aligned}\label{eq:hom}
    R&=\sqrt{1-\frac{2\Delta}{K\cos(\alpha)}}\\
    \Omega&=\overline \omega -\tan(\alpha)(K\cos(\alpha)-\Delta)\;.
\end{aligned}
\end{equation}
These equations show that our choice of the mean frequency $\overline \omega=\tan(\alpha)(K\cos(\alpha)-\Delta)$
defines a co-rotating frame in which the homogeneous coherent solution is a steady state rather than a limit-cycle (i.e., $\Omega=0$).
This simplifies the analytical calculations that follow.
\\

Next we focus on the stability of the coherent homogeneous state $Z>0$ given by Eq.~\eqref{eq:hom}.
The same analysis for the incoherent homogeneous solution $Z=0$ is provided in Appendix~\ref{appIcoherence}, where we show
that no instabilities occur below the Kuramoto synchronization transition $\text{H}_0$.
Let us consider a small perturbation of the coherent homogeneous state, $Z_\sigma=Z+z_\sigma$.
Linearizing Eq.~\eqref{eq:oaM} we obtain
\begin{equation}
\begin{aligned}\label{eq:linearization}
        \dot z_\sigma &= (-\Delta + i \overline \omega -  K  R^2 e^{i\alpha}) z_\sigma \\
    &+ \frac{K }{2}\sum_{\kappa=1}^M c_{\sigma \kappa} (z_\kappa e^{-i\alpha} - R^2 z^*_\kappa e^{i\alpha} )\;.\\
\end{aligned}
\end{equation}

In order to simplify the notation, in the following steps we express complex quantities in vectorial form: 
$z = (x , y)^T$ where $x$ and $y$ are the real and imaginary parts of $z$, respectively.
Then, Eq.~\eqref{eq:linearization} can be expressed as
\begin{equation}\label{eq:linmat}
        \dot z_\sigma=
        A z_\sigma + B\sum_{\kappa=1}^M c_{\sigma \kappa} z_\kappa,
\end{equation}
where
\begin{equation}
A=   \begin{pmatrix}
    \Delta-K\cos(\alpha) & -\Delta \tan(\alpha) \\[10pt]
    \Delta \tan(\alpha) & \Delta -K \cos(\alpha)
    \end{pmatrix}
\end{equation}
and
\begin{equation}
B= \begin{pmatrix}
    \Delta & \Delta \tan(\alpha)\\[10pt]
    -K\sin(\alpha)+\Delta \tan(\alpha) & K\cos(\alpha)-\Delta
    \end{pmatrix}\;
\end{equation}
These expressions for $A$ and $B$ are obtained by inserting the values of $\overline \omega$ and $R$ derived from Eq.~\eqref{eq:hom} into Eq.~\eqref{eq:linearization}, and separating real and imaginary parts.

Next we apply the master stability formalism (MSF)~\cite{pecora1998}, in order to simplify the analysis of Eq.~\eqref{eq:linmat}\footnote{
See Refs.~[\onlinecite{arenas2008,porter2016,Ashwin2016}] for introductory reviews on the MSF formalism.
Although most studies on the MSF rely on diffusive coupling or variations of it via a Laplacian matrix (see, e.g., [\onlinecite{Nakao2014}]),
the method is applicable
to any setup having a homogeneous solution (see, e.g., [\onlinecite{clusella2023}]).
}.
This method consists on decomposing the perturbation $z_\sigma$ on the basis provided
by the diagonalization of the connectivity matrix $C$.
We denote the eigenvalues of the connectivity matrix as $\Lambda_1 \geq \Lambda_2 \geq \cdots \geq \Lambda_N $,
and the corresponding eigenvectors as $\boldsymbol{\Psi}^{(k)}=(\Psi_1^{(k)},\dots,\Psi_N^{(k)})^T$ with $k=1,\dots,N$.
Therefore
\begin{equation}\label{eq:eig}
 C \boldsymbol{\Psi}^{(k)} = \Lambda_k \boldsymbol{\Psi}^{(k)}\;.
\end{equation}
We now express the perturbation $z_\sigma$ on the basis $\boldsymbol{\Psi}^{(k)}$ as
\begin{equation}\label{eq:change}
z_\sigma = \sum_{k=1}^N u_k \otimes \Psi_\sigma^{(k)} ,
\end{equation}
where $u_k\in\mathbb{C}$ are the coordinates of $z_\sigma$ in the new basis and $\otimes$ is the
Kronecker product.

Inserting Eq.~\eqref{eq:change} in the linearization provided by Eq.~\eqref{eq:linearization}
and making use of Eq.~\eqref{eq:eig}
we derive the following expression:
\begin{equation}
\begin{aligned}
\dot z_\sigma &= \sum_{k=1}^N \dot u_k \otimes\Psi_\sigma^{(k)}\\
&= A \sum_{k=1}^N u_k \otimes\Psi_\sigma^{(k)} + B \sum_{\kappa=1}^N c_{\sigma \kappa}\sum_{k=1}^N u_k \otimes\Psi_\kappa^{(k)}\\
&=\sum_{k=1}^N \left( A  u_k \otimes\Psi_\sigma^{(k)} + B u_k\otimes \sum_{\kappa=1}^N c_{\sigma \kappa}  \Psi_\kappa^{(k)}\right) \\
&=\sum_{k=1}^N \left( A  u_k \otimes\Psi_\sigma^{(k)} + \Lambda_k B u_k\otimes \Psi_\sigma^{(k)}\right)\\
&=\sum_{k=1}^N \left( A + \Lambda_kB  \right) u_k \otimes\Psi_\sigma^{(k)}.
\end{aligned}
\end{equation}
Since the eigenvectors $\boldsymbol{\Psi}^{(k)}$ are a basis of $\mathbb{R}^N$, 
linear independence provides
\begin{equation}
\dot u_k =  \left( A + \Lambda_k B \right)  u_k\;.
\end{equation}
Therefore, we decomposed the $2N\times 2N$ linear system Eq.~\eqref{eq:linmat} 
into $N$ 2-dimensional linear systems that depend on the eigenvalues $\Lambda_k$.

The eigenvalues $\lambda_k^{(\pm)}$ of the matrices $A+\Lambda_k B$ for $k=1,\dots,N$ characterize the
stability of the homogeneous solution~\eqref{eq:hom}. 
In fact, they correspond to the \emph{Floquet exponents} of the periodic solution $Z=Re^{i\Omega t}$.
The analytical expression for the eigenvalues reads:
\begin{equation}\label{eq:eigs}
\begin{aligned}
\lambda_k^{(\pm)}=&\Delta +K\cos(\alpha)\left(\frac{\Lambda_k}{2} -1 \right) \\
\pm&\Biggl\{ \frac{K^2\Lambda_k^2\cos^2(\alpha)}{4} \\
&+ K \Lambda_k \Delta \cos(\alpha) \left[ \tan^2(\alpha)(1-\Lambda_k)-\Lambda_k \right] \\
&+\Delta^2\left[(\Lambda_k^2-1)\tan^2(\alpha)+\Lambda_k^2)\right]
\Biggr\}^{1/2}\;. 
\end{aligned}
\end{equation}
This corresponds to a dispersion relation analogous to that in reaction-diffusion systems, 
with the structural eigenmodes $\boldsymbol{\Psi}^{(k)}$ playing the role of wave functions.

The eigenvalues $\lambda_k^{(\pm)}$ depend explicitly on the parameters of the system ($\Delta$, $\alpha$, and $K$), as
well as on the structural eigenvalues $\Lambda_k$.
Using the row-normalization of $C$ and the Gershgorin circle theorem\cite{wilkinson1967} we 
can assert that $|\Lambda_k|\leq 1$. Moreover, 
the largest structural eigenvalue is $\Lambda_1=1$, and the corresponding
eigenvector corresponds to a uniform perturbation, i.e.,
\begin{equation}
    \boldsymbol{\Psi}^{(1)} = \frac{1}{\sqrt{N}}(1,\cdots,1)^T.
\end{equation}
From the expression of $\lambda_k$ in Eq.~\eqref{eq:eigs} we see that $\lambda_1^{+}=0$,
as it should, since the homogeneous state is a limit-cycle in a co-rotating frame. 
The eigenvalues $\lambda_k$ for $k>1$ indicate the growth rates of perturbations that
are transverse to the homogeneous manifold.

For the case $\Delta=0$, the expression of the system eigenvalues simplifies to $\lambda_k^{(+)} = K\cos(\alpha)(\Lambda_k -1)$
and $\lambda^{(-)}_k=-K\cos(\alpha)$.
Since the coherent state emerges for $K\cos(\alpha)>2\Delta=0$,
this proves that, in networks without disorder,
the homogeneous state is always stable, consistent with the fact
that chimera states in these systems always coexist with the fully synchronized solution~\cite{abrams2008}.

\begin{figure}[t]
  \centerline{\includegraphics[width=0.5\textwidth]{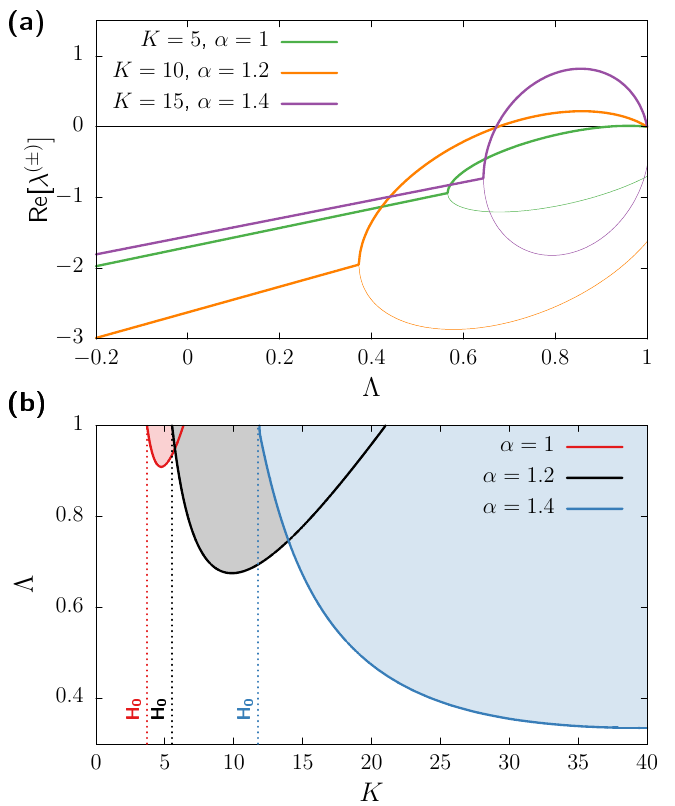}}
        \caption{Transverse instabilities of homogeneous states in networks of Kuramoto-Sakaguchi populations Eq.~\eqref{eq:oaM}.
        (a) Real part of the eigenvalues (Floquet exponents) characterizing the stability of the homogeneous solution $\lambda^{(\pm)}$ as a function of the structural matrix eigenvalues $\Lambda$. Thick and thin curves correspond to $\lambda^{(+)}$ and $\lambda^{(-)}$
        as given by the master stability function in Eq.~\eqref{eq:eigs} respectively.
        Green, orange, and purple curves correspond to $(K,\alpha)=(5,1)$,
        $(10,1.2)$, and $(15,1.4)$.
        (b) Color-shaded regions show the regions of transverse instabilities in the $(K,\Lambda)$ plane,
        with solid lines showing the loci of the bifurcation ($\text{Re}[\lambda^{(+)}] = 0$ in Eq.~\eqref{eq:eigs}). 
        Dotted vertical lines indicate the Kuramoto synchronization transition $\text{H}_0$ 
        at which the coherent homogeneous solution emerges. Results shown for $\alpha=1$ (red), 1.2 (black), and 1.4 (blue). }
  \label{fig:msf}
\end{figure}

For $\Delta>0$, however, the scenario changes. In this case it is possible to
perform a parameter reduction (see Appendix~\ref{app:param}). Setting $\Delta = 1$ without loss of generality, we analyze the system for different values of $K$ and $\alpha$.
Figure~\ref{fig:msf}(a) shows $\Real[\lambda^{(\pm)}]$ as a function of $\Lambda$
for different parameter sets. 
For the selected values, the dispersion relation displays a positive region,
indicating the existence of transverse instabilities,
which always emerge through real eigenvalues $\lambda^{(+)}\in\mathbb{R}$
\footnote{
The homogeneous synchronized solution emerges only for $\Delta-K\cos(\alpha)/2<0$, 
which leads to $\Delta +K\cos(\alpha)\left(\frac{\Lambda}{2} -1 \right)<0$ in Eq.~\eqref{eq:eigs} (since $\Delta\geq 0$ and $-1\leq \Lambda\leq 1$).
Therefore,  eigenvalues with positive real part in Eq.~\eqref{eq:eigs}
always correspond to real eigenvalues. Complex conjugate eigenvalues
might exist, but they correspond to stable directions.
}.
Overall, the shape of the dispersion relation
resembles that of the Benjamin-Feir instability in the complex Ginzburg-Landau equation~\cite{kuramoto1984}.

Setting $\Real[\lambda^{(+)}]=0$ in Eq.~\eqref{eq:eigs} determines the regions where transverse
instabilities of the homogeneous state arise in parameter space.
Figure~\ref{fig:msf}(b) shows such regions for three different values of $\alpha$ in the $(K,\Lambda)$-plane.
A small region of instabilities arise already for $\alpha=1$, and widens as $\alpha\to \pi/2$, the critical value at which the homogeneous coherent solution vanishes through H$_0$.
This is consistent with previous findings showing that chimera and other symmetry-breaking states emerge
only for $\alpha$ close to, but below, $\pi/2$\cite{abrams2004,abrams2008,kotwal2017}.
In the next sections we review the dynamics that emerge in this region in the two-population model,
ring networks, and Erdős-Rényi connectivites. For this analysis we fix $\alpha=1.2$, a value
not that close to the H$_0$ bifurcation, and still displaying a large instability region.

\section{Review of the two-population model}

\begin{figure*}[t]
  \centerline{\includegraphics[width=\textwidth]{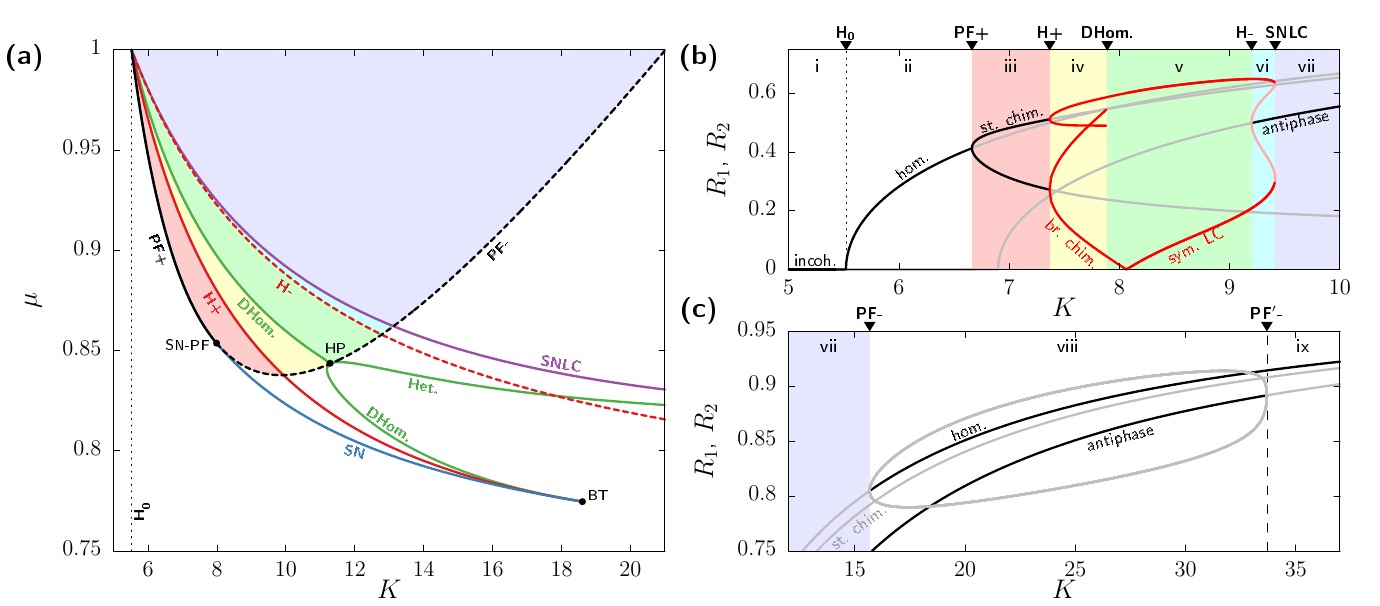}}
        \caption{Bifurcation diagrams of the two population model (Eq.~\eqref{eq:oaM} with Eq.~\eqref{eq:W2}) for $\alpha=1.2$ and $\Delta=1$.
        (a) Two-parameter bifurcation diagram. Solid lines indicate supercritical bifurcations (+) 
        and dashed lines indicate subcritical bifurcations (-).
        Black curves correspond to the pitchfork bifurcation (PF) of the homogeneous state (given by Eq.~\eqref{eq:eigs}),
        red curves indicate Hopf bifurcations (H),
        green curves indicate double homoclinic (DHom.) and heteroclinic (Het.) bifurcations,
        the blue curve indicates a saddle-node bifurcation (SN),
        and the purple curve indicates a saddle-node of limit-cycles (SNLC).
        Straight dotted thin line indicates the Kuramoto synchronization transition (H$_0$).
        Color-shaded regions indicate instability of the homogeneous solution,
        with each color corresponding to a different attractor set:
        pink for stable steady chimeras,
        yellow for breathing chimeras,
        green for symmetric oscillations,
        blue for bistability between symmetric oscillations and antiphase state,
        and purple for stability of the antiphase fixed point only.
        (b,c) One-parameter bifurcation diagram for $\mu=0.9$.
        Solid curves with different colors indicate different attractor types:
        black for stable fixed points,
        grey for unstable fixed points,
        red for stable limit-cycles,
        and light red for unstable limit-cycles.
        Relevant bifurcations are labelled on the top of the figures, with
        the different regions between bifurcations shaded with the same color scheme as in panel (a).
        Homogeneous and antiphase solutions together with their corresponding bifurcations (H$_0$,PF,H-) have been obtained analytically.
        All other results have been obtained with auto-07p\cite{auto07p} (see Appendix~\ref{app:simulations}).
        }
  \label{fig:2popsdiag}
\end{figure*}

As a first example we revisit the two-population model.
Previous works on this setup
showed the emergence of chimera and other states from a pitchfork bifurcation of the homogeneous synchronized state if $\Delta>0$\cite{laing2009,laing2012,kotwal2017,guo2021}.
Here we show that such bifurcation corresponds to a transverse instability provided by Eq.~\eqref{eq:eigs}, and review 
some of the dynamics emerging from this symmetry-breaking.
As in the previous section, we fix $\Delta=1$ without loss of generality (see Appendix \ref{app:param}).

The connectivity matrix in the 2-population model reads
\begin{equation}\label{eq:W2}
C = 
\begin{pmatrix} 
\mu & 1-\mu\\[10pt]
1-\mu & \mu
\end{pmatrix}\;.
\end{equation}
where $\mu\in[0,1]$ is a parameter controlling the coupling strength within the population,
and $1-\mu$ corresponds to the coupling strength across the two populations.

\begin{figure*}[th]
  \centerline{\includegraphics[width=\textwidth]{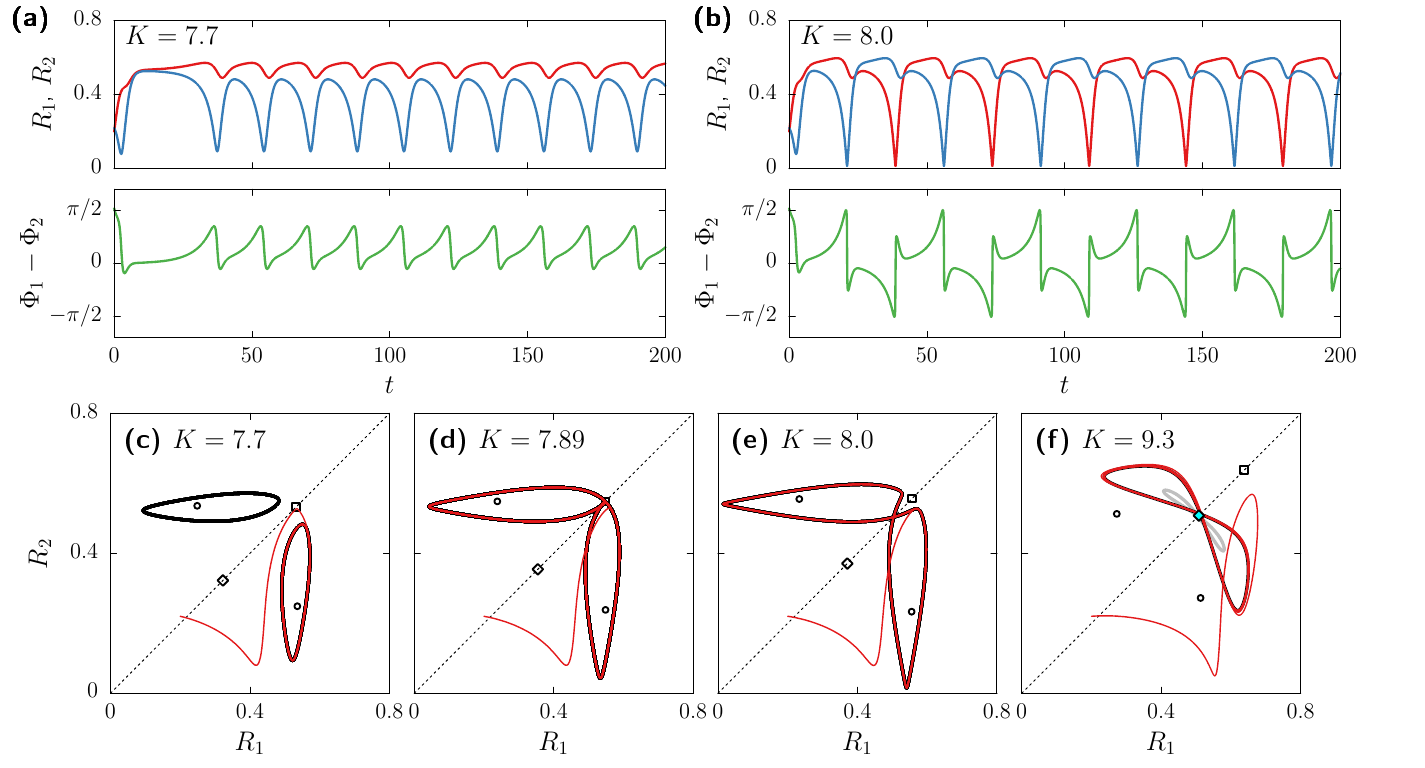}}
        \caption{Oscillatory dynamics in the 2-population Kuramoto model (Eq.~\eqref{eq:oaM} with Eq.~\eqref{eq:W2}) for $\mu=0.9$, $\alpha=1.2$, and $\Delta=1$.
        (a,b) Time series of the Kuramoto order parameters and phase difference of the two populations at a breathing chimera state (panel (a), $K=7.7$) and asymmetric limit-cycle (panel (c), $K=8$).
        (c-f) Phase portraits for different values of $K$ showing breathing chimera (panel (c)),
        the symmetric limit-cycle (panels (d) and (e)), and bistability
        between the antiphase state and the symmetric limit-cycle (panel (f)).
        Bold black curves
        show the stable limit-cycles and red curves show an exemplary trajectory.
        Grey curven in panel (f) shows the unstable limit-cycle.
        Open symbols correspond to the different unstable fixed points: 
        square ({\tiny $\square$}) indicates the homogeneous state, 
        diamond ($\diamond$) indicates the antiphase state,
        circles ($\circ$) indicate the two steady chimera states.
        Blue-colored diamond in panel (f) indicates the stable antiphase state.
        }
  \label{fig:breathing}
\end{figure*}

The eigenvalues of this connectivity matrix read $\Lambda_1=1$ and $\Lambda_2=2\mu-1$.
Substituting $\Lambda_2$ in Eq.~\eqref{eq:eigs} and setting $\text{Re}[\lambda_2^{(+)}]=0$ provides 
an analytical expression for the pitchfork bifurcation of the homogeneous state as an implicit equation on the system parameters.
The solid black curve in Figure~\ref{fig:2popsdiag}(a) shows such bifurcation line in the $(K,\mu)$-plane,
with the color-shaded regions above the line indicating the areas where the homogeneous state is unstable (see also black curve in Fig.~\ref{fig:msf}(b)).
Within this region, four different states might exist (apart from the incoherent and coherent homogeneous states)\footnote{For $\alpha=1.2$ we did not find the chaotic state unveiled in Ref.~[\onlinecite{laing2012}], but it can be attained,
for instance, at $(K,\mu,\alpha,\Delta)=(20 , 0.9 ,1.25, 1)$}:
\begin{itemize}
\item {\bf Steady chimera} (pink region in Fig.~\ref{fig:2popsdiag}a), a fixed point characterized by  $R_1\neq  R_2$.
\item {\bf Breathing chimera} (yellow region in Fig.~\ref{fig:2popsdiag}a), a limit-cycle characterized by one population having always a larger degree of synchrony than the other, i.e., either $R_1(t) > R_2(t)$ or $R_2(t)> R_1(t)$ for all $t>0$ (see Fig.~\ref{fig:breathing}(a)).
\item {\bf Symmetric limit-cycle} (green and blue regions in Fig.~\ref{fig:2popsdiag}a), characterized by $R_1(t)=R_2(t+T/2)$, where $T$ is the period
of the oscillation (see Fig.~\ref{fig:breathing}(b)).
\item {\bf Antiphase fixed point} (blue and purple regions in Fig.~\ref{fig:2popsdiag}a), characterized by $R_1(t)=R_2(t)$ but $\Phi_1-\Phi_2=\pi$. A thorough analysis 
of this state, including its linear stability, is provided in Appendix~\ref{appAntiphase}.
\end{itemize}

In order to illustrate the emergence and nature of these different states,
we first fix $\mu=0.9$ and study the system dynamics upon increasing $K$ with the help of numerical continuation software\cite{auto07p}.
The resulting bifurcation diagram is depicted in Figs.~\ref{fig:2popsdiag}(b,c),
and can be split in nine different regions (labelled i-ix) delimited by different bifurcations:
\begin{enumerate}[label=(\roman*)]
	\item For $K<K_c\approx 5.52$ the incoherent state $Z_\sigma  = 0$ is the only stable solution.
		This state loses stability at the Kuramoto synchronization transition H$_0$, leading to the emergence of the coherent homogeneous state.
   
    \item The coherent homogeneous state is the global attractor of the system from H$_0$ up to $K\approx 6.66$, when it loses stability through the 
    supercritical pitchfork bifurcation (PF+) given by Eq.~\eqref{eq:eigs}.
   
    \item Above PF+, two stable fixed points emerge, corresponding to steady chimeras (black lines in Fig.~\ref{fig:2popsdiag}(b), red-shaded region). Although the steady chimeras coexist with the unstable homogeneous and antiphase fixed points, they are the sole attractor in this region of the parameter space.
   
    \item At $K\approx 7.37$ the steady chimeras undergo a supercritical Hopf bifurcation (H+) leading to the emergence
    of two stable limit-cycles, corresponding to the breathing chimera (red lines in Fig.~\ref{fig:2popsdiag}(b), yellow-shaded region). 
    Figure~\ref{fig:breathing}(a) shows exemplary time series of this state, and Fig.~\ref{fig:breathing}(c)
    the corresponding phase portrait, with black curves depicting the limit-cycles and the red curve showing an example of a trajectory.
    
    \item The breathing chimera limit-cycles eventually collide with the unstable coherent homogeneous state at $K\approx 7.89$, merging in a single limit-cycle through a double homoclinic bifurcation (Dhom, also known as \emph{gluing bifurcation}~\cite{lopez2000,pazo2001,laing2012}). The resulting attractor corresponds to a symmetric limit cycle in which the two populations alternate the same level of synchrony half a period apart (red lines in Fig~\ref{fig:2popsdiag}(b), green-shaded region).
    Figure~\ref{fig:breathing}(b) shows exemplary time series of this regime and Figs.~\ref{fig:breathing}(d,e) 
    depict two phase portraits, with $K$ in panel (d) very close to the gluing bifurcation.
    
    \item At $K\approx 9.19$, the antiphase state of the system, which is unstable for smaller values of $K$, 
    becomes stable through a subcritical Hopf bifurcation (H-, see Appendix~\ref{appAntiphase}).
    Therefore, in this region of the bifurcation diagram there is bistability between the antiphase state and the symmetric limit-cycle (black and red lines in Fig.~\ref{fig:2popsdiag}(b), blue-shaded region).
    Figure~\ref{fig:breathing}(f) shows the phase portrait of this bistable regime, with the unstable limit-cycle represented
    by a grey curve. 
    
    \item The symmetric limit-cycle collides with the unstable limit-cycle at $K\approx 9.41$, thus both solutions vanish through a saddle-node of limit-cycles (SNLC).
    Then, the antiphase state becomes the only attractor for a wide region of the parameter space  (black curve in Figs.~\ref{fig:2popsdiag}(b,c), purple-shaded region).
    
    \item At $K\approx 15.67$, the homogeneous fixed point recovers stability through the (subcritical) pitchfork bifurcation
    provided by Eq.~\eqref{eq:eigs} (see PF- in Fig.~\ref{fig:2popsdiag}(c)).
    Therefore, in this region both homogeneous and antiphase states share stability (black curves in Fig.~\ref{fig:2popsdiag}(c)).
    
    \item Finally, the unstable fixed points emerging from PF- collide with the antiphase solution at a new subcritical pitchfork bifurcation 
    (PF'-) at $K\approx 33.69$ (see Appendix~\ref{appAntiphase} for an analytical derivation of PF'-).
    From this point on, further increase of $K$ does not lead to any other bifurcation, and
    the homogeneous state remains the only (global) attractor of the system  (black curve in Fig.~\ref{fig:2popsdiag}(c)).
\end{enumerate}

The two-parameter bifurcation diagram in Fig.~\ref{fig:2popsdiag}(a) shows that the bifurcation
scenario we have described for $\mu=0.9$ holds for the entire region of instability of the homogeneous
state. The diagram also shows the loci of a saddle node bifurcation (SN) that
joints the pitchfork bifurcation (PF) at a fold-pitchfork codimension-2 bifurcation (SN-PF).
At this codimension 2 point, the symmetry breaking bifurcation turns from supercritical to subcritical\cite{laing2009,kotwal2017}.
This SN branch limits a region of bistability between the homogeneous fixed point and stable chimeras.
Within this region, following a route similar to the one described before, the stable chimera turns into a breathing state at
the supercritical Hopf (H+). Nonetheless,
in this case the gluing bifurcation leads to the disappearance of limit-cycle solutions, since the homogeneous state is stable. 
Finally, there is an additional region of bistability between the asymmetric limit cycle
and the homogeneous state bounded by a heteroclinic bifurcation and the saddle-node of limit cycles (SNLC).

Altogether, this analysis unveils a rich dynamical landscape emerging from the pitchfork bifurcation that
breaks the symmetry of the homogeneous state.
These results complement those presented in previous works
for the same system~\cite{laing2009,laing2012,kotwal2017,guo2021} but without relying on weak heterogeneity.
Moreover, the pitchfork bifurcation of the homogeneous state (PF) and the bifurcations from the antiphase state (H- and PF'-) have been obtained analytically as implicit equations on the system parameters.

\section{Ring topology}

\begin{figure}[t]
  \centerline{\includegraphics[width=0.5\textwidth]{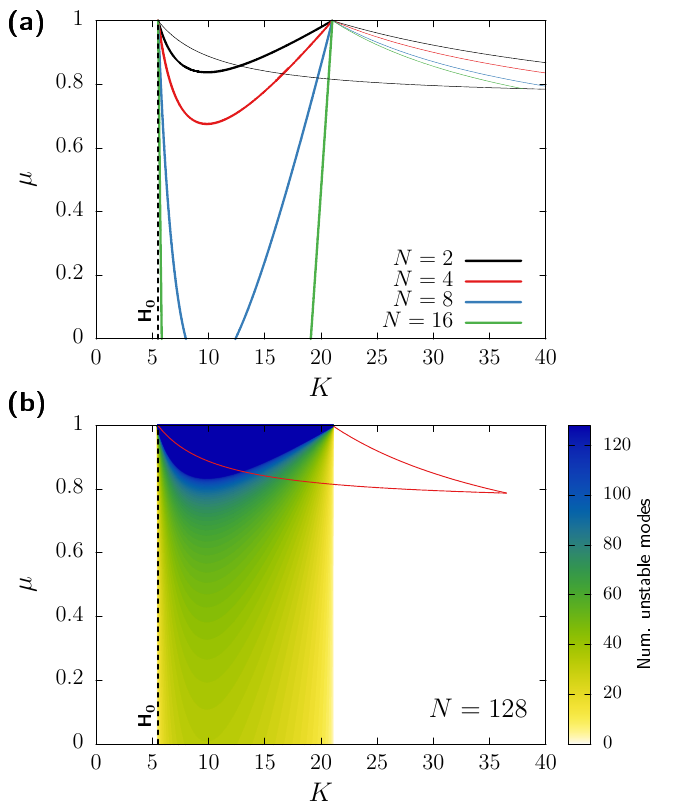}}
        \caption{Transverse instabilities of the homogeneous and antiphase state in ring networks ($\alpha=1.2$ and $\Delta=1$).
        (a) Thick lines indicate the region of transverse instability of the homogeneous state for different $N$.
        Thin curves indicate the region of stability of the antiphase state.
        (b) Color-shaded regions show the number of unstable modes of the homogeneous state for a network of $N=128$ populations. Red curve indicates the corresponding region of stability of the antiphase state.
        Vertical black dashed line indicates the Kuramoto synchronization (H$_0$) transition in both panels.}
  \label{fig:ringdiagram}
\end{figure}

Now we consider a ring network composed of $N>2$ populations connected via nearest neighbors.
The connecitivity matrix $C=(c_{\sigma\kappa})$ reads
\begin{equation}\label{eq:ring}
c_{\sigma\kappa}=
\begin{cases}
\mu &\text{if}\quad \sigma=\kappa\\
\frac{1-\mu}{2} & \text{if}\quad |\sigma-\kappa|=1\text{ or }N-1\\
0 &\text{otherwise}\;.
\end{cases}
\end{equation}

This scenario was studied in Ref.~[\onlinecite{lee2023}] for the case of $N=6$ oscillators, where
they unveiled different forms of stable and (non-chaotic) breathing chimera states for heterogeneous populations. 
Also, Ref.~[\onlinecite{laing2023}] analyzed a similar ring model but with non-local coupling, which displayed chaotic
solutions that vanish upon increasing system size. 
Here we show that, in our setup, transverse instabilities of the homogeneous state
lead to chaotic dynamics that show an extensive character with $N$, i.e., the 
dynamical complexity of the system scales linearly with the system size.
 
Since $C$ is a circulant matrix, its eigenvalues have a simple expression.
In order to preserve the ordering from larger to smaller, we write them as
\begin{equation}\label{eq:lambdaring}
    \Lambda_k = 
   \begin{cases}
        \displaystyle{\mu + (1-\mu)\cos\left(\frac{\pi}{N} (k-1)\right)} & \text{if }k\text{ is odd},\\[10pt]
        \displaystyle{\mu + (1-\mu)\cos\left(\frac{\pi}{N}k\right)} & \text{if }k\text{ is even},
    \end{cases}
\end{equation}
for $k=1,\dots,N$. 
The corresponding eigenvectors are the discrete Fourier modes
\begin{equation}\label{eq:ringvectors}
    \Psi^{(k)}_j = \begin{cases}
    \displaystyle{\frac{e^{\frac{\pi j}{N}(k-1)i}}{\sqrt{N}}}& \text{if }k\text{ is odd},\\[10pt]
    \displaystyle{\frac{e^{- \frac{\pi j}{N}ki}}{\sqrt{N}}}& \text{if }k\text{ is even},
    \end{cases}
\end{equation}
where $k=1,\dots,N$. 

As $N$ increases, the second largest structural eigenvalue $\Lambda_2$ tends to 1.
Therefore, from Fig.~\ref{fig:msf}(b), we infer that the region of transverse instability expands with $N$.
Figure~\ref{fig:ringdiagram}(a) illustrates this situation:
Thick continuous curves depict the bifurcation obtained in the $(K,\mu)$-plane by setting $\text{Re}[\lambda^{(+)}]=0$ in Eq.~\eqref{eq:eigs}.
As $N$ increases, the region of instability becomes larger,
covering a wide band of $K$ values irrespective of $\mu$ if $N$ is sufficiently large. 
Moreover, the emergence of instabilities occurs arbitrarily close to the Kuramoto synchronization transition (H$_0$, 
vertical black dashed line in Fig.~\ref{fig:ringdiagram}(a)).

Another important aspect of the ring spectrum in Eq.~\eqref{eq:lambdaring} is that it is delocalized, i.e., it covers the interval $\Lambda_k\in[2\mu-1,1]$ densely as $N\to\infty$.
Therefore,
several structural eigenmodes $\Lambda_k$ might concurrently correspond to different unstable
directions.
For instance, the colored region in Fig.~\ref{fig:ringdiagram}(b) shows
the number of unstable directions of the homogeneous state in the $(K,\mu)$-plane
for $N=128$. The number of positive growth rates generally increases with $\mu$,
with a large region (dark blue) where all $\lambda_k^{(+)}>0$ for $k>1$.
In fact, this region coincides with the region of transverse instability for the two-population model, since the last mode turning unstable corresponds to $\Lambda_N=2\mu -1$.
\begin{figure*}[t]
  \centerline{\includegraphics[width=\textwidth]{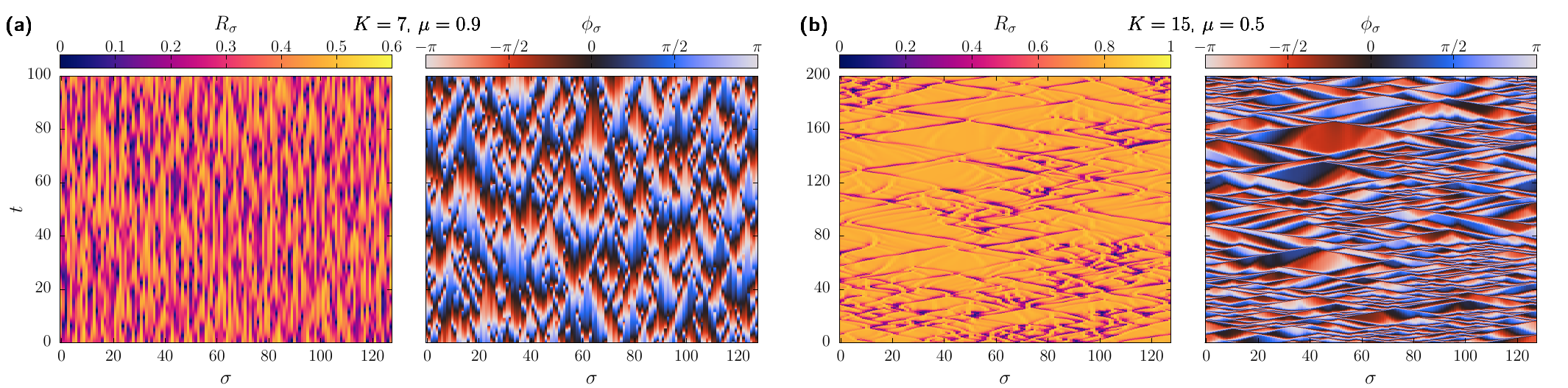}}
        \caption{Chaotic dynamics of the ring model for different parameter values ($\alpha=1.2$ and $\Delta=1$).
        (a) Kuramoto order parameter of each population $R_\sigma$ and corresponding phase $\Phi_\sigma$
        for $K=7$ and $\mu=0.9$ after a transient of 1000 time units.
        (b) Same quantities for $K=15$ and $\mu=0.5$.}
  \label{fig:patterns}
\end{figure*}

The regularity of ring topologies also allows for the existence of antiphase and other twisted states~\cite{lee2023}.
Appendix~\ref{appAntiphase} provides a linear stability analysis for antiphase states in ring networks.
The thin lines in Fig.~\ref{fig:ringdiagram}(a) and the red line in Fig.~\ref{fig:ringdiagram}(b) delimit the region of stable antiphase dynamics.
This region shrinks by increasing $N$, although the changes are less prominent
than those of the homogeneous state (see Appendix~\ref{appAntiphase} for a further discussion).

Overall, in large ring networks, we expect to have complex dynamics in wide regions of parameter space
due to the expansion of the transverse instability region and the number of unstable directions of the homogeneous state.
Figure~\ref{fig:patterns} shows two instances of these complex dynamics for $N=128$. 
Pannels (a) depict $R_\sigma$ and $\Phi_\sigma$ of each population over time
for parameters in the region where the number of unstable directions of the homogeneous state is maximal ($K=7$, $\mu=0.9$).
The dynamics of $R_\sigma$ are rather disordered, with $\phi_\sigma$ showing some indications
of spatial structure. 
Instead, panels (b) correspond to parameters associated with less unstable directions ($K=15$, $\mu=0.5$).
In this case,
the dynamics of both $R_\sigma$ and $\Phi_\sigma$ evolve more regularly, in spite of their seemingly chaotic behavior. 
These qualitative observations posit the existence of space-time chaos emerging from the transverse instabilities reported here.
In the rest of this section, we assess the chaotic and complex behavior of the system by means of numerical Lyapunov analysis.
Appendix~\ref{app:measures} contains precise definitions of the monitored quantities.

\begin{figure*}[t]
  \centerline{\includegraphics[width=1\textwidth]{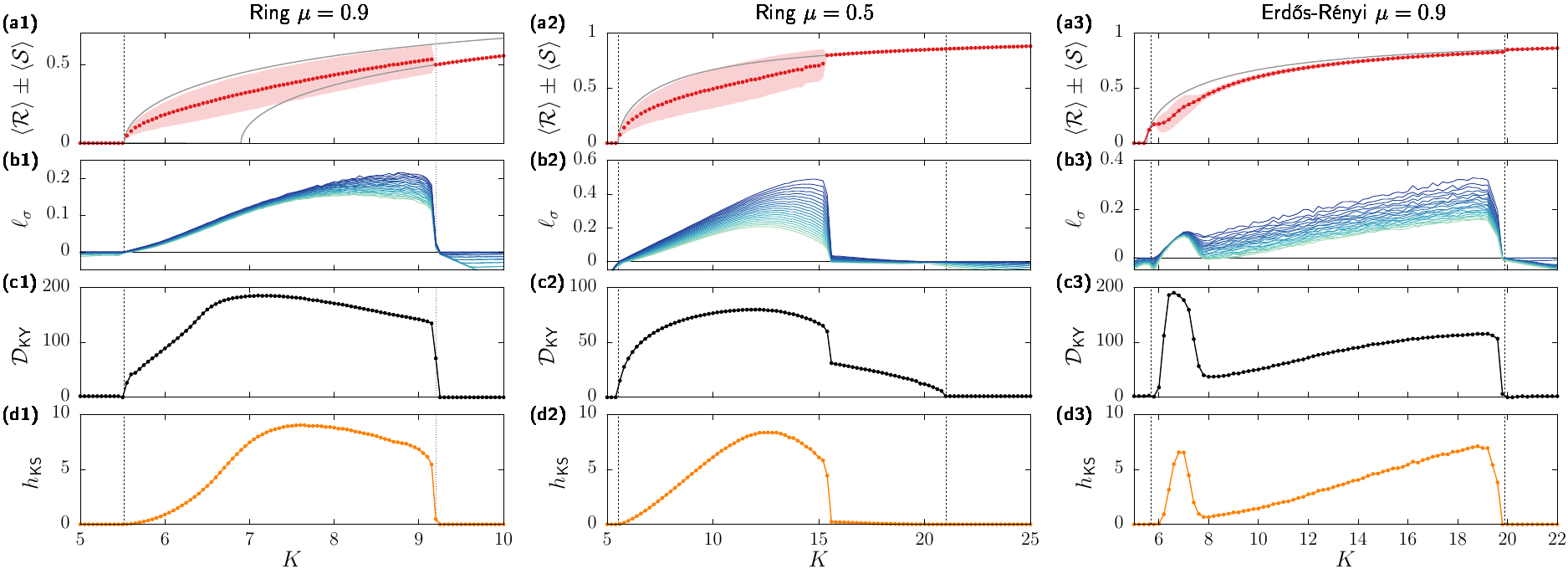}}
        \caption{
        Simulations of ring (columns 1 and 2) and Erdős-Rényi (column 3) networks with $N=128$ ($\alpha=1.2$ and $\Delta=1$).
        (a) Red circles indicate time average mean activity $\langle \mathcal{R} \rangle$,
        with the red-shaded region showing the corresponding
        level of heterogeneity $\langle \mathcal{R}\rangle \pm \langle \mathcal{S}\rangle $.
        Grey curves correspond to the value of $R$ for the homogeneous states given by Eq.~\eqref{eq:hom}.
        In panel (a1) an additional grey curve shows the antiphase state (Eq.~\eqref{eq:Rhat}).
        (b) 20 largest Lyapunov exponents $\ell_\sigma$.
        (c) Kaplan-Yorke fractal dimension computed using the (complete) Lyapunov spectra
        as given by Eq.~\eqref{eq:kaplanyorke}.
        (d) Kolmogorov-Sinai dynamical entropy computed from Eq.~\eqref{eq:entropy}.
        In all panels, vertical dashed lines indicate the emergence and disappearance of transverse instabilities as given by Eq.~\eqref{eq:eigs}.
        Dotted vertical line in panels (a1-d1) indicates the emergence of stable antiphase state as given by Eq.~\eqref{eq:eigs_antiphase}. }
  \label{fig:K}
\end{figure*}

First, we focus on a fixed network size ($N=128$) and study the dynamics by varying $K$.
Figures~\ref{fig:K}(a1-d1) show the numerical results obtained with fixed $\mu=0.9$,
thus covering the region of maximal number of unstabilities.
The red circles in Fig.~\ref{fig:K}(a1) show the time average of the network's mean Kuramoto order parameter, $\mathcal{R}$.
The shaded region shows the corresponding standard deviation $\mathcal{S}$ (see Appendix~\ref{app:measures} for specific definitions).
The grey continuous lines indicate the value of $\mathcal{R}$ corresponding to the homogeneous and antiphase states (Eqs.~\eqref{eq:hom} and~\eqref{eq:Rhat}). 
As predicted, the nontrivial dynamics emerge just after the Kuramoto transition to synchrony H$_0$, 
and appear to be the only attracting state until the stabilization of the antiphase state (see dotted vertical lines). 

To characterize the complexity of these non-trivial states, Fig.~\ref{fig:K}(b1)
displays the 20 largest Lyapunov exponents $\ell_\sigma$ of the network. 
At the transverse instability, many exponents become positive at once,
thus indicating the emergence of high-dimensional chaotic dynamics.
Indeed, the Kaplan-Yorke formula (Fig.~\ref{fig:K}(c1)) indicates
a fractal dimension with about $\mathcal{D}_\text{KY}\approx 28$ degrees of freedom
just after the transition.
The attractor dimension increases smoothly until it reaches a maximum of $\mathcal{D}_\text{KY}\approx 200$,
and then slowly decreases until vanishing at the stabilization of the antiphase state.
Similarly, the Kolmogorov-Sinai dynamical entropy $h_\text{KS}$ (Fig.~\ref{fig:K}(d1)) also increases starting at the transverse instability,
albeit it does so less abruptly. This measure indicates that the maximum complexity of the
trajectories is reached around $K \approx 7.5$.
Altogether, these results establish the existence of high-dimensional chaos
arising from the transverse instabilities of the homogeneous state for $\mu=0.9$.

Next, we switch to $\mu=0.5$, for which the homogeneous state has less unstable directions.
Moreover, the antiphase state does not exist for this value of $\mu$, and therefore, we expect
non-trivial dynamics for a larger $K$-range (note the difference in the x-axis range between columns 1 and 2 in Fig.~\ref{fig:K}).
Figures~\ref{fig:K}(a2-d2) show that the scenario starts off analogously to that of $\mu=0.9$:
at the transverse instability (see vertical dotted line) the system transitions from incoherence to a state with a large degree of heterogeneity.
Also, several Lyapunov exponents become positive at once,
resulting into both $\mathcal{D}_\text{KY}$ and $h_\text{KS}$ increasing rapidly.
Again, this corresponds to a case of space-time chaos.

Around $K\approx 15.8$ a transition occurs and the irregularity
of the system drops drastically.
The dynamics emerging from this bifurcation, which is not detected by our theory, proved challenging to characterize.
First, the system suddenly evolves very close to the homogeneous state. 
In fact, our measure of network heterogeneity $\mathcal{S}$ is too small to be visually appreciated above the transition 
in Fig.~\ref{fig:K}(a2) (see pink shaded region).
Nonetheless, a close inspection of the simulations (not shown) revealed irregular patterns in both $R_\sigma$ and $\Phi_\sigma$.
Second, such spatiotemporal dynamics evolve slowly compared to the states encountered so far, requiring
longer simulations (see Appendix~\ref{app:simulations}).
Such slow time-scales are mainly reflected by the magnitude of the positive Lyapunov exponents in Fig.~\ref{fig:K}(b2)
and the drop on dynamical entropy shown in Fig.~\ref{fig:K}(d2). 
In spite of these peculiarities, the number of positive Lyapunov exponents, and thus the value of $\mathcal{D}_\text{KY}$ for $15.8 \lesssim K\lesssim 21$ (see Fig.~\ref{fig:K}(c3)) indicate that this corresponds to a high-dimensional chaotic regime
\footnote{We also performed simulations with adiabatic increase and decrease of $K$ in other to test
for potential bistability between the fast and slow chaotic states. The outcome consistently
showed the disappearance of the faster dynamics around $K\approx $.}. 
 Finally, around $K\approx 21.6$ the dynamics 
 fall back to the homogeneous state, as predicted by the stability analysis (second vertical dashed line in the middle-row panels of Fig.~\ref{fig:K}).

\begin{figure*}[t]
  \centerline{\includegraphics[width=1\textwidth]{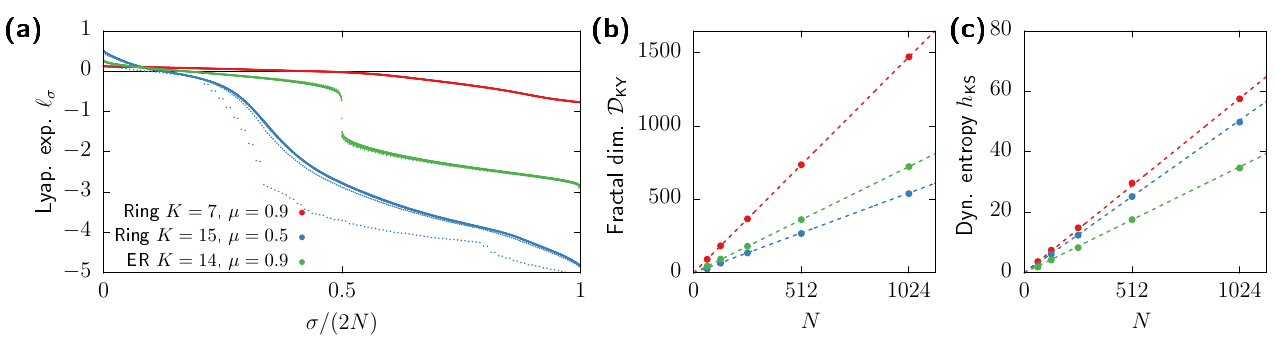}}
        \caption{Extensivity of chaos in ring and Erdős-Rényi (ER) networks ($\alpha=1.2$ and $\Delta=1$).
        (a) Complete Lyapunov spectra for different network sizes for the simulations of the ring and ER networks.
        For each parameter set, dots indicate the results for networks with $N=64,128,256, 512$ and 1024 populations.
        Nonetheless, in most cases differences across system sizes are barely visible.
        (b,c) Fractal dimension $D_\text{KY}$ (panel b) and dynamical entropy (panel c) $h_\text{KS}$ 
        computed for different system sizes $N$ in three different simulation setups.
        Circles correspond to numerical simulations. Dashed lines correspond to a linear regression.
        In all these plots, red corresponds to ring networks with $K=7$ and $\mu=0.9$,
        blue corresponds to ring networks with $K=15$ and $\mu=0.5$,
        and green corresponds to ER networks with $K=14$ and $\mu=0.9$.
        }
  \label{fig:extensive}
\end{figure*}

So far, the numerical exploration for $\mu=0.9$ and $\mu=0.5$ proved the existence of
highly complex spatiotemporal regimes emerging from transverse instabilities 
of the homogeneous state in rings of $N=128$ populations.
Next, we test the robustness and scaling these results upon varying the system size $N$.
Figure~\ref{fig:extensive} shows results from simulations of rings composed of increased number $N$ of oscillators (red and blue symbols)
for selected parameter values.
Panel (a) shows the Lyapunov spectra computed for different system sizes with node index normalized by the system size $\sigma/(2N)$.
For the tested parameter values ($(K,\mu)=(7,0.9)$, red symbols, and $(15,0.5)$, blue symbols)
the spectra show a good collapse upon increasing $N$, i.e., the Lyapunov exponents
converge to a well defined profile.
The only exception of the smaller networks for the second set (blue dots).
These robust spectra for increasing $N$ indicate 
extensive chaos, i.e. a chaotic regime whose complexity scales linearly with the system size. 
To further explore this possibility, Figs.~\ref{fig:extensive}(b) and (c) show 
the computed fractal dimension and dynamical entropy for the same parameter values upon increasing $N$.
In all cases, an accurate scaling  $\mathcal{D}_\text{KY}\propto N$ and $h_\text{KS}\propto N$ is observed.

Overall, these numerical results indicate that transverse instabilities of a homogeneous state
lead to extensive chaos in ring networks of oscillator populations with nearest neighbour coupling. 
In the next section we investigate the case of irregular topologies.

\section{Erdős-Rényi networks}

\begin{figure}[t]
  \centerline{\includegraphics[width=0.5\textwidth]{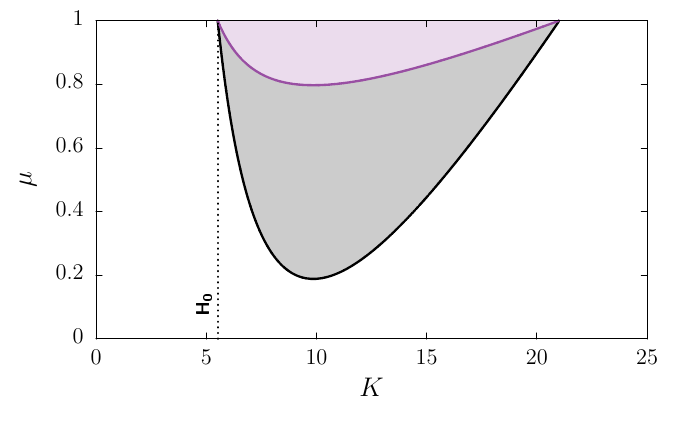}}
        \caption{Approximated region of transverse instabilities of the homogeneous state in Erdős-Rényi networks
        with average degree $\overline d = 10$.
        The grey-shaded region indicates where at least one mode is unstable (i.e., $\text{Re}[\lambda_2^{(+)}]>0$).
        The purple-shaded region indicates where $N-1$ modes are unstable. 
        The vertical dashed line indicates the Kuramoto synchronization transition H$_0$.
        }
  \label{fig:ERdiagram}
\end{figure}

Here we discuss the dynamics that emerge in irregular network topologies.
In order to simplify the analysis, we restrict ourselves on
connectivities derived from  Erdős-Rényi (ER) networks with fixed average degree $\overline d =10$. 
Let $A=(a_{\sigma \kappa})$ be the adjacency matrix of an (undirected) Erdős-Rényi network (i.e., $a_{\sigma \kappa}\in\{1,0\}$).
Then, we consider a connectivity matrix $C=(c_{\sigma \kappa})$ given by
\begin{equation}
c_{\sigma \kappa}=
\begin{cases} 
\mu\quad  &\text{ if }\quad \sigma =\kappa\\
\displaystyle{\frac{1-\mu}{d_\sigma}}a_{\sigma \kappa}\quad&\text{ if }\quad\sigma \neq \kappa,
\end{cases}
\end{equation}
where $d_\sigma$ is the degree of node $\sigma$ in the Erdős-Rényi model.
This connectivity preserves row-normalization, as required for the existence of 
homogeneous states, as well as introduces the self-coupling strength parameter $\mu$.

Unlike the ring scenario, we are unaware of explicit analytical results on the spectra
of $C$ (which does not need to coincide with the spectra of the adjacency matrix $A$). 
However, combining analytical arguments with a numerical exploration, we can provide an accurate estimation of the region of transverse instabilities.
By the Gershgoring circle theorem~\cite{wilkinson1967}, the eigenvalues of $C$ are contained
in the interval $\Lambda_k\in[1, 2\mu-1]$.
Therefore, based on Fig.~\ref{fig:msf}(b), transverse instabilities of the homogeneous state
will occur. The shape of the region will depend on the second largest eigenvalue $\Lambda_2$.
If $\tilde \Lambda $ is an eigenvalue of the matrix $D^{-1}A = (a_{\sigma \kappa}/d_\sigma)$,
then $\Lambda = \mu + (1-\mu)\tilde \Lambda$ is an eigenvalue of $C$.
Averages over $100$ simulated networks with different sizes (not shown) indicate that $\tilde \Lambda$ are densely distributed
in the interval [-0.6, 0.6]. In particular, $\tilde \Lambda_2 \approx 0.6$ as $N$ increases, 
and therefore $\Lambda_2 \approx 0.4\mu + 0.6$.
Using this approximation in the MSF, Eq.~\eqref{eq:eigs}, provides the region of transverse
instabilities for ER networks in the $(K,\mu)$ parameter space.
The grey-shadowed region in Fig.~\ref{fig:ERdiagram} shows the region where at least
one mode of the homogeneous state is unstable. 
Also in this case, the number of unstable directions in the region increases with $N$, 
with the purple region in Fig.~\ref{fig:ERdiagram} indicating the parameter
values for which the homogeneous state has a maximum number of unstable modes ($N-1$).

In order to capture the emergent dynamics, Figure~\ref{fig:K}(a3-d3) displays
the results of simulations performed with fixed $\mu=0.9$ and different values of $K$ for an ER network of $N=128$ populations.
As in the ring networks, the numerical outcome shows a scenario of consistent
high-dimensional chaos for most of the transverse instability region (indicated by vertical dashed lines).
Interestingly, two different types of complex dynamics seem to emerge:
First, the level of heterogeneity captured by $\mathcal{S}$ shows a
region of high dynamical variety among nodes (see red-shaded region in Fig.~\ref{fig:K}(a3), $6 \lesssim  K \lesssim 8$). 
In this case the Lyapunov spectra shows many exponents with similar positive values (panel b3),
which translates into a very high Kaplan-Yorke dimension (panel c3) and dynamical entropy (panel d4).
At $K\approx 8$, all measures show a significant reduction of complexity,
indicating a transition to a different chaotic regime from $8 \lesssim K\lesssim 20$.
In this case, the patterns lose heterogeneity,
but the Lyapunov spectra still shows a regime of high-dimensional chaos.
In fact, the number and value of the largest exponents increase monotonically with $K$,
leading to an equally gradual increase of $\mathcal{D}_\text{KY}$ and $h_\text{KS}$
until the homogeneous state becomes stable again (at $K\approx 20$).

As in the ring model, we now test the robustness of the chaotic dynamics upon increasing network size. 
The green dots in Figure~\ref{fig:extensive}(a)
show the entire Lyapunov spectra for $K=15$ and $\mu=0.9$ upon increasing system size.
Overall, the different spectra collapse to a single well-defined curve.
Moreover,
the green symbols in Fig.~\ref{fig:extensive}(b,c)
indicate that both the fractal dimension and the dynamical entropy increase linearly with $N$,
confirming thus the extensive nature of the chaotic regime in ER topologies.

\section{Conclusions}

Recent works have studied the dynamics of networks of identical Kuramoto-Sakaguchi populations 
by means of the Ott-Antonsen equations under
differerent parameter setups, coupling schemes, and system sizes\cite{abrams2008,laing2009,laing2009a,laing2012,martens2010,martens2010a,lee2023,lee2023a,bick2020}.
This paper provides a leap forward in these efforts by providing
a general expression for the instabilities of the homogeneous state 
as a function of the eigenmodes of the structural connectivity matrix.
We proved that, in networks without disorder ($\Delta = 0$), the homogeneous state is always transversally stable regardless
the topology.
On the other hand, for disordered populations ($\Delta>0$), transverse instabilities occur.
These instabilities, which lead to chimera states in two-population networks\cite{laing2009,kotwal2017,guo2021},
give rise to extensive chaos in large nearest-neighbour ring and random connectivity schemes.

The emergence of extensive chaos in ring networks could apparently conflict with the findings 
of Laing~\cite{laing2023}, which studied a similar setup but with non-local connectivity via a discrete kernel.
In their work, the row-normalization of the connectivity matrix is preserved, thus
our theoretical framework also applies.
The eigenvalues of their connectivity matrix can be analytically computed, and we found they are all zero except for three: 
$\Lambda_1=1$, $\Lambda_2=\Lambda_3>0$. In particular, $\lim_{N\to\infty} \Lambda_2 = B/2$ where $B=0.35$.
This localized spectra differs largely from the spectra of nearest neighbour ring topologies given by Eq.~\eqref{eq:lambdaring},
which covers the interval $[2\mu-1,1]$ densely for large $N$.
Although a more in-detail comparison between both setups is out of the scope of this paper,
it is likely that the transverse instabilitites reported in Ref.~[\onlinecite{laing2023}] only
occur through $\Lambda_2$ and $\Lambda_3$, rather through than several concurrently unstable eigenmodes.
This might explain the disappearance of chaos for large networks in that setup.

These differences between non-local and local interactions in rings indicate that network topology and, specifically,
their spectra, play an important role on the properties of the resulting dynamical states.
Here we restricted ourselves on simple structures, without an in-depth analysis of
how specific topological features affect the emergent behavior.
Future work should analyze the effect of varying network density in both ring and Erdős-Rényi networks,
and the impact of other topologies such as scale-free or small world.

On a different note, the extensive nature of the chaotic states and the patterns displayed in ring networks
are an indication that this form of space-time chaos could correspond to a turbulent state.
Interestingly, the dependence of the growth rates on the structural eigenmodes depicted in Fig.~\ref{fig:msf}(a) 
resembles that of the Benjamin-Feir
instability in the complex Ginzburg-Landau equation\cite{Benjamin1967}. 
Similar turbulent states have been studied in arrays of individual Kuramoto-Sakaguchi oscillators with non-local
coupling~\cite{wolfrum2016,bolotov2018,bolotov2021}.
Further analysis should thus address to what extent the properties of our setup for increasing $N$ match those of phase or amplitude turbulence 
in spatially extended systems~\cite{kuramoto1984}.

Here we have studied the dynamics of the Kuramoto-Sakaguchi populations
at the thermodynamic limit via the Ott-Antonsen equations.
The global attractiveness of this low-dimensional manifold for $\Delta>0$ has been recently proven\cite{cestnik2022},
strengthening the validity of this approach to study networks of Kuramoto-Sakaguchi populations.
Nonetheless, previous works have shown that  (low-dimensional) chaotic dynamics unveiled by the Ott-Antonsen equations
might display different properties in finite-size simulations~\cite{laing2012}.
A detailed comparison of simulations of the microscopic system~\eqref{eq:general_kuramoto} with finite $M$,
and those of the Ott-Antonsen description Eq.~\eqref{eq:oaM}, could 
provide valuable insights to understanding the convergence of Kuramoto-Sakaguchi ensembles
to the low-dimensional manifold.
Nonetheless, such study would
require a large computational effort, due to the large number of degrees of freedom involved.

Finally, although our study does not stem from any particular physical setup, the Kuramoto-Sakaguchi model successfully describes a wide variety of real-world phenomena, including power grids\cite{filatrella2008}, circadian rhythms\cite{strogatz1987}, and bacterial biofilms\cite{liu2017}.
A particularly explored application is neural dynamics, where the Kuramoto-Sakaguchi model has been rigorously derived from spiking neuron models\cite{clusella2022a}.
In this context, the mean-field description provided by the Ott-Antonsen equations leads to a simplified neural mass model\cite{bick2020}.
Thus, our analysis of the coupled system Eq.~\eqref{eq:oaM} provides a simple, yet grounded, framework to
analyze whole brain dynamics where each brain region is represented by a Kuramoto-Sakaguchi population.
In fact, the study of transverse instabilities in whole brain models is a powerful tool to unveil the multiscale dynamics of these systems,
even when the single-node dynamics evolve according to generic neural mass models\cite{clusella2023}.

\section{Acknowledgements}

The authors would like to thank Diego Pazó and Alessandro Torcini for useful feedback on this work.
PC and JGO have received funding from the Future and Emerging Technologies Programme (FET) of the European Union’s Horizon 2020 research and innovation programme (project NEUROTWIN, grant agreement No 101017716). JGO was also financially supported by the Spanish Ministry of Science and Innovation, the Spanish State Research Agency and FEDER (project reference PID2021-127311NB-I00), and by the ICREA Academia program. 
This work has done while PF was a student at the ETSETB, Universitat Politècnica de Catalunya.

\appendix 

\section{Measures of collective activity}\label{app:measures}

Here we describe the different quantitative measures employed to characterize 
the dynamics emerging from transverse instabilities in large networks of oscillator populations.

\begin{itemize}
\item To capture the macroscopic behavior of the system, we compute the mean Kuramoto order parameter
\begin{equation}
\mathcal{R}(t):=\frac{1}{N}\sum_{\sigma = 1}^N R_\sigma(t)
\end{equation}
and its instantaneous standard deviation on the network
\begin{equation}
\mathcal{S}(t) = \frac{1}{\sqrt{N}}\left\{\sum_{\sigma = 1 }^N \left[R_\sigma(t) - \mathcal{R}(t) \right] ^2\right\}^\frac{1}{2}.
\end{equation}
If the network is at a heterogeneous state (in $R_\sigma$), then $ \mathcal{S} > 0$.

\item The chaotic activity of a system can be characterized by the corresponding Lyapunov exponents~\cite{Politi2013,Pikovsky2016} which we denote by $\ell_\sigma$,
where $\ell_1\geq \ell_2\geq \cdots\geq \ell_N$.
We compute the Lyapunov exponents using the standard dynamical algorithm based on QR-decomposition~\cite{benettin1980,Pikovsky2016}.
In particular, simulations have been performed using the \emph{DynamicalSystems.jl} Julia package~\cite{Datseris2018} (see Appendix \ref{app:simulations} for more details on numerical simulations).

\item The fractal dimension of a chaotic attractor can be approximated using the Kaplan-Yorke formula~\cite{Kaplan1979}, given by
\begin{equation}\label{eq:kaplanyorke}
    \mathcal{D}_\text{KY}=j + \frac{\sum_{i=1}^j \ell_i}{|\ell_{j+1}|}
\end{equation}
where $j$ is such that $\sum_{i=1}^j \ell_i\geq 0$ but $\sum_{i_1}^{j+1} \ell_i<0$.

\item Another measure of dynamical complexity is the Kolmogorov-Sinai dynamical entropy $h_\text{KS}$,
which measures the growth rate of distinguishable trajectories in phase space~\cite{Pikovsky2016,sinai2009}. 
Following Pesin's formula, $h_\text{KS}$ can be computed as the sum of all positive Lyapunov exponents
\begin{equation}\label{eq:entropy}
    h_\text{KS}=\sum_{i=1}^n \ell_i.
\end{equation}
where $n$ is such that $\ell_n\geq 0$ and $\ell_{n+1}<0$.
\end{itemize}

\section{Numerical simulations and numerical continuation}\label{app:simulations}

The code used in this paper is openly available at [\onlinecite{repository}] (also accessible through \href{https://github.com/pclus/KuramotoPopulationNetwork}{https://github.com/pclus/KuramotoPopulationNetwork}).
Simulations of the model and computation of Lyapunov exponents have been performed
in Julia via the DynamicalSystems.jl package~\cite{Datseris2018}.
In particular we have employed the Runge-Kutta 4 algorithm with a fixed time step of $dt=10^{-2}$.

Simulations of the system employed an initial transient evolution of $6000$ time units.
This is a conservative transient time set to achieve a good convergence
near the bifurcations, but for most parameter values shorter transients provide the same results.
After this transient, we simulate both phase space and tangent space dynamics for the computation
of the Lyapunov exponents.
The tangent space is initially evolved for 200 time units, after which we start computing all the Lyapunov exponents
for a total of 1000 time units.
The QR decomposition for computation of the spectra is invoked at every time unit (i.e. $1/dt$ time steps).
The computation of the Kaplan-Yorke formula requires a tolerance to discern zero and non-zero exponents, which we set to $10^{-4}$.

The results depicted in Fig.~\ref{fig:K}(a2-d2)
require longer computation times due to the existence of the slow chaotic regime.
In that case, the integration time step was chosen to be $dt=10^{-1}$, the transient in tangent space consisted of $10^4$ time units, 
and the Lyapunov exponents were computed for $3\cdot 10^4$ time units with the QR decomposition at every 10 time units. 
Also in this case, we lowered the tolerance for the Kaplan-Yorke formula to $2\cdot 10^{-5}$.

Simulations of the ring network for $\mu=0.9$ were initialized
at the antiphase state with an additional small random perturbation.
This prevents the system from stalling into a long chaotic transient in the region where the antiphase state is stable.
For all other cases, since the antiphase state does not exist or is unstable, simulations were initialized at the homogeneous
state with a random small perturbation. 

The bifurcation diagrams of Fig.~\ref{fig:2popsdiag} were partially obtained with the numerical
continuation software auto-07p\cite{auto07p}.
In order to simplify the analysis, the 2-population model with the connectivity matrix~\eqref{eq:W2}
was rewritten as a three-dimensional system in real space by
considering the polar representation of the Kuramoto
complex parameters $Z_1=R_1e^{i\Phi_1}$ and $Z_2=R_2e^{i\Phi_2}$, and defining a new variable $\Theta=\Phi_1-\Phi_2$. Then, Eq.~\eqref{eq:oaM} reads
\begin{equation}\label{eq:oa2simplified}
\begin{aligned}
\dot R_1 &= -\Delta R_1 \\&+ K\left(\frac{1-R^2_1}{2}\right) \Bigl[\mu R_1\cos{(\alpha)}  + (1-\mu) R_2\cos{(\Theta-\alpha)}\Bigr] \\
\dot R_2 &= -\Delta R_2 \\&+ K\left(\frac{1-R^2_2}{2}\right) \Bigl[\mu R_2\cos{(\alpha)}  + (1-\mu) R_1\cos{(\Theta+\alpha)}\Bigr] \\
\dot \Theta &= K\left(\frac{1+R_1^2}{2R_1}\right)\left[ \mu R_1 \sin(\alpha)-(1-\mu)R_2\sin(\Theta-\alpha) \right] \\
&- K\left(\frac{1+R_2^2}{2R_2}\right)\left[ \mu R_2 \sin(\alpha)+(1-\mu)R_1\sin(\Theta+\alpha) \right]\;.
\end{aligned}
\end{equation}
A tutorial guide to obtain some of the bifurcations using the numerical continuation capabilities of auto-07p is openly available in the published repository\cite{repository}.

\section{Parameter reduction}\label{app:param}

System~\eqref{eq:oaM} contains three parameters: $\Delta$, $K$, and $\alpha$.
It is possible to rescale this system in order to reduce the parameter number to two.
Some works\cite{laing2009,kotwal2017,guo2021} consider systems in which time is rescaled as $\tilde t=Kt$, so that the coupling strength $K$
can be removed from the equations or, equivalently, let it constant as $K=1$.
In this case Eq.~\eqref{eq:oaM} reads
\begin{equation}
    \frac{\dd Z_\sigma}{\dd\tilde t}  = -D Z_\sigma + 
    \frac{1}{2}\sum_{\kappa=1}^N c_{\sigma \kappa} \left( Z_\kappa e^{-i\alpha}  - Z^2_\sigma Z^*_\kappa e^{i\alpha} \right)
\end{equation}
where $D = \Delta/K$ and $K>0$.

Another option is to rescale time as $\tilde t= \Delta t$, such that~\eqref{eq:oaM} reads 
\begin{equation}
    \frac{\dd Z_\sigma}{\dd\tilde t}   = - Z_\sigma + 
    \frac{\tilde K}{2}\sum_{\kappa=1}^N c_{\sigma \kappa} \left( Z_\kappa e^{-i\alpha}  - Z^2_\sigma Z^*_\kappa e^{i\alpha} \right).
\end{equation}
with $\tilde K=K/\Delta$ and $\Delta > 0$.
This would be equivalent to let $\Delta=1$ fixed in the equations.

In this work we choose not to rescale any parameter in the system, so the analysis remains valid in the limit cases $\Delta = 0$ and $K=0$.
Nonetheless, we are mostly interested in the heterogeneous case ($\Delta>0$), thus we let $\Delta=1$ without loss of generality.
Therefore, in the 2-population model, the equivalence between the formulation of previous studies~\cite{laing2009,laing2012,kotwal2017,guo2021} and that given in Eq. \eqref{eq:oaM} with $\Delta=1$ is simply $D=1/K$.

\section{Antiphase states in ring networks}\label{appAntiphase}

Here we provide the stablity analysis of 
antiphase states in topologies given by either Eq.\eqref{eq:W2} or Eq.~\eqref{eq:ring} with an even number of populations $N$.
This analysis is performed in close analogy to that performed in Section~\ref{section:homogeneous}.
Antiphase states are solutions of Eq.~\eqref{eq:oaM} given by
\begin{equation}
    Z_\sigma = 
    \begin{cases}
    \displaystyle{\hat Re^{i\hat \Omega t}}&\text{ if }\sigma\text{ is  even}\\[10pt]
    \displaystyle{\hat Re^{i(\hat \Omega t +\pi)}}&\text{ if }\sigma\text{ is  odd}.\\
    \end{cases}
\end{equation}
Inserting these expressions into Eq.~\eqref{eq:oaM} and using the corresponding expression of $c_{\sigma \kappa}$
we obtain that 
\begin{equation}
    \begin{aligned}\label{eq:Rhat}
    \hat R = \sqrt{1-\frac{2\Delta}{K(2\mu-1)\cos(\alpha)}}\quad\text{and}\\
    \hat \Omega = \overline\omega - \tan(\alpha)[ K(2\mu-1)\cos(\alpha) - \Delta].
    \end{aligned}
\end{equation}
By linearizing~\eqref{eq:oaM} around this solution we obtain
\begin{equation}
\begin{aligned}
        \dot z_\sigma &= (-\Delta + i \overline \omega -  K(2\mu-1)  \hat R^2 e^{i\alpha}) z_\sigma \\
    &+ \frac{K }{2}\sum_{\kappa=1}^M c_{\sigma \kappa} (z_\kappa e^{-i\alpha} - \hat R^2 z^*_\kappa e^{i\alpha} )\;.\\
\end{aligned}
\end{equation}
By choosing $\overline \omega = \tan(\alpha)[ K(2\mu-1)\cos(\alpha) - \Delta]$ and replacing the expression of $\hat R$ from Eq.~\eqref{eq:Rhat} 
we obtain
\begin{equation}
        \dot z_\sigma=
        \hat A z_\sigma + \hat B\sum_{\kappa=1}^N c_{\sigma \kappa} z_\kappa
\end{equation}
where
\begin{equation}
\hat A=   \begin{pmatrix}
    \Delta - K \cos(\alpha) (2\mu-1) & -\Delta \tan(\alpha) \\[10pt]
    \Delta \tan(\alpha) & \Delta - K \cos(\alpha) (2\mu-1)
    \end{pmatrix}
\end{equation}
and
\begin{equation}
\hat B= \begin{pmatrix}
    \frac{\Delta}{2\mu-1} & \frac{\Delta}{2\mu-1} \tan(\alpha)\\[10pt]
    -K\sin(\alpha)+\frac{\Delta}{2\mu-1} \tan(\alpha) & K\cos(\alpha)-\frac{\Delta}{2\mu-1}
    \end{pmatrix}\;.
\end{equation} 
Analogously to the stability of the homogeneous states shown in Section~\ref{section:homogeneous},
we apply the MSF formalism based on decomposing a perturbation on the basis of the eigenvectors of 
the connectivity matrix $C$. As a result we obtain that the stability of the
antiphase state is given by the eigenvalues of the $N$ $2\times 2$ matrices:
\begin{equation}
    \hat M_k =\hat A + \Lambda_k\hat B\quad\text{for}\quad k=1,\dots,N;
\end{equation}
where $\Lambda_k$ are obtained from Eq.~\eqref{eq:lambdaring}.
The explicit expression of the system eigenvalues read:
\begin{equation}\label{eq:eigs_antiphase}
\begin{aligned}
\hat \lambda_k^{(\pm)}=&\Delta +K\cos(\alpha)\left(\frac{\Lambda_k}{2} +1-2\mu \right) \\
\pm&\Biggl\{ \frac{K^2\Lambda_k^2\cos^2(\alpha)}{4} \\
&+ \frac{K \Lambda_k \Delta \cos(\alpha)}{2\mu-1} \left[ \tan^2(\alpha)(2\mu-1-\Lambda_k)-\Lambda_k \right] \\
&+\frac{\Delta^2}{(2\mu-1)^2}\left[(\Lambda_k^2-(2\mu-1)^2)\tan^2(\alpha)+\Lambda_k^2)\right]
\Biggr\}^{1/2}\;. 
\end{aligned}
\end{equation}
From this expression we shall highlight several aspects:
\begin{itemize}
\item As in the homogeneous state, the antiphase state is a periodic solution
in a co-rotating frame. Therefore at least one of the eigenvalues has to be 0.
This corresponds to $\Lambda_{N}=2\mu-1$, as the associated
eigenvector indicates perturbations acting opposite on each node consecutively,
i.e., from Eq.~\eqref{eq:ringvectors}:
\begin{equation}
    \Psi^{(N)}_j = \begin{cases}
    \frac{1}{\sqrt{N}}& \text{if }j\text{ is odd},\\[10pt]
    \frac{-1}{\sqrt{N}}& \text{if }j\text{ is even}.
    \end{cases}
\end{equation}
\item Taking into account the previous point, for $N=2$ the stability of 
the antiphase solution is controlled by the eigenvalue associated to $k=1$. 
Substituting thus $\Lambda_1=1$ in Eq.~\eqref{eq:eigs_antiphase} and solving for $\lambda_1^{(\pm)}=0$
we obtain the bifurcations from the antiphase state in the 2-population model.
This eigenvalue can be zero through a pair of complex conjugates or a single real eigenvalue,
defining thus the subcritical Hopf (H-) and the pitchfork (PF'-) bifurcations displayed in Fig.~\ref{fig:2popsdiag}.

\item For $N>2$ other eigenvalues rather than $\Lambda_1$ can lead to instabilities of the antiphase state.
This is the case of Fig.~\ref{fig:ringdiagram}(a), in which the left branch of the instability is always given by
$\Lambda_1$, but the right branch corresponds to other modes depending on $N$.
\end{itemize}

\section{Transverse instabilities of the incoherent state}\label{appIcoherence}

Here study the transverse instabilities of the incoherent homogeneous state $ Z_\sigma = 0$ for $\sigma=1,\dots,N$.
Our aim is to show that transverse instabilities do not arise when this state is stable, i.e., when $0< K\cos(\alpha) < 2\Delta $.
Without loss of generality, we impose $\overline \omega =0$.
Then, by linearizing Eq.~\eqref{eq:oaM} around the incoherent solution ($Z_\sigma =0$) we obtain
\begin{equation}
\begin{aligned}
        \dot z_\sigma &= -\Delta  \omega z_\sigma + \frac{K }{2}\sum_{\kappa=1}^M c_{\sigma \kappa} z_\kappa e^{-i\alpha} \\
        &= A^{(0)} z_\sigma + B^{(0)} \sum_{\kappa=1}^N c_{\sigma \kappa} z_\kappa
        \;,
\end{aligned}
\end{equation}
where
\begin{equation}
A^{(0)} =   \begin{pmatrix}
    -\Delta  & 0 \\[10pt]
    0 & -\Delta 
    \end{pmatrix}
\end{equation}
and
\begin{equation}
B^{(0)} = \frac{K}{2}\begin{pmatrix}
    \cos(\alpha) & \sin(\alpha)\\[10pt]
    -\sin(\alpha) & \cos(\alpha)
    \end{pmatrix}\;.
\end{equation}

Following again the master stability formalism presented in Section \ref{section:homogeneous},
the stability of this solution is controlled by the eigenvalues $\lambda_k^{(\pm)}$ of the matrices
\begin{equation}
     M^{(0)}_k = A^{(0)} + \Lambda_k B^{(0)} \quad\text{for}\quad k=1,\dots,N;
\end{equation}
Now, instead of finding a general expression for the eigenvalues it is enough to see that
\begin{equation}
\begin{aligned}
\text{Tr}(M^{(0)}_k) = \lambda_k^{(+)} + \lambda_k^{(-)} = \Lambda_k K \cos(\alpha) - 2\Delta\;.
\end{aligned}
\end{equation}
The incoherent solution $Z^{(0)}$ is stable within the homogeneous manifold Eq.~\eqref{eq:hommanifold}
if and only if $K\cos(\alpha) < 2\Delta $. In this case, since $-1<\Lambda_k<1$ and $\Delta>0$, we have that $\text{Tr}(M^{(0)}_k)<0$.
Therefore, transverse instabilities of the incoherent state may emerge only if this state is already
unstable to the homogeneous perturbation.

\bibliography{references}

\end{document}